\documentclass[%
 reprint, %add a p here
 amsmath,amssymb,
 aps,
prl,
]{revtex4-2}
\usepackage{graphicx}% Include figure files
\usepackage{dcolumn}% Align table columns on decimal point
\usepackage{bm}% bold math
\usepackage{amsmath, amssymb, braket, multirow, booktabs, array, cancel} %I ADDED THESE ONES
\usepackage{xcolor}
\usepackage[normalem]{ulem}
\begin{document}

% \preprint{APS/123-QED}

\title{\textbf{Experimental measurement of quantum-first-passage-time distributions} 
}% 

\author{Joseph M. Ryan}
\email{Contact author: joseph.ryan@duke.edu}
\author{Simon Gorbaty}
\author{Thomas J. Kessler}
\author{Mitchell G. Peaks}
\author{Stephen W. Teitsworth}
\author{Crystal Noel}
% \author{Simon Gorbaty}%
%  \email{Contact author: Second.Author@institution.edu}
\affiliation{Duke Quantum Center, Department of Electrical and Computer Engineering and Department of Physics, Duke University, Durham, NC 27708, USA}%

\date{\today}% It is always \today, today,

\begin{abstract}
Classical First-Passage-Time Distributions (FPTDs) have been extensively studied both theoretically and experimentally. Their quantum counterparts—Quantum First-Passage-Time Distributions (QFPTDs)—remain largely unexplored and have deep implications for both fundamental physics and the development of emerging quantum technologies. We measure the first QFPTDs using a motional mode of a single trapped ion.  We develop a novel composite-phase laser pulse sequence to perform tunable stroboscopic \textcolor{black}{single-shot} projective measurements of the motional state of a trapped ion.  We measure QFPTDs of the ion energy when coupled to electric-field noise.  
The measurement protocol developed here is broadly applicable to other quantum systems and provides a powerful method for exploring a broad range of QFPTD phenomena.  
With these results we open a new field of experimental investigations of QFPT processes with potential future relevance to quantum search algorithms, unraveling connections between classical and quantum dynamics, and study of the quantum measurement problem. 
\end{abstract}

\maketitle

\textit{Introduction}---The first-passage time is typically defined as the first time that a system dynamical observable is measured to be outside a \textit{surviving domain}.  Distributions of these times, known as First-Passage-Time Distributions (FPTDs), have a long history in science. They reveal individual trajectory dynamics that ensemble quantities \textcolor{black}{may} fail to capture~\cite{sensing_advantage_2, sensinv_advantage_1, Redner_2001}, as exemplified by Schr{\"o}dinger's clarification of Millikan's oil drop experiment~\cite{Schroedinger_FPT}. FPTDs have since been used in diverse areas such as in the study of activation in chemical and biological processes~\cite{Redner_2001, Hanggi_RevModPhys.62.251, activation_1, activation_2, activation_3}, current-switching in electronic transport structures~\cite{Teitsworth_PhysRevLett.109.026801, Teitsworth_EPJB_2019}, economics and market models~\cite{Market_FPT_1, Market_FPT_2, Market_FPT_3, Market_FPT_4}, and climate science~\cite{Stechmann_2014, Lucente_2022, Tesfay_2025}.  

\begin{figure}[t]
    \centering
    \includegraphics[width=1\linewidth]{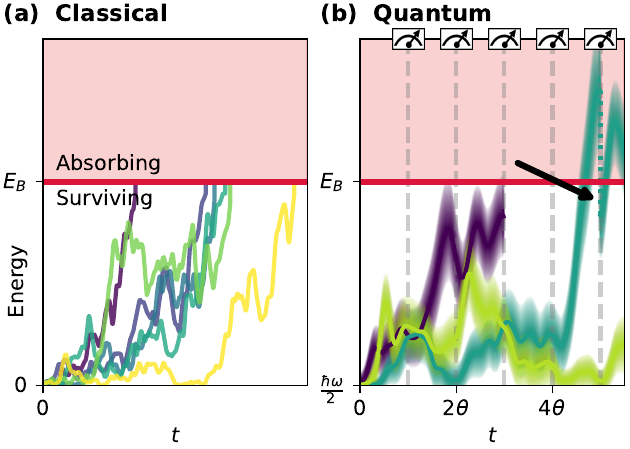}
    \caption{Conceptual comparison of quantum and classical FPTs with energy threshold $E_B$. (a) Sample classical FPT trajectories with continuous measurement, which end when the energy is greater than $E_B$.  \textcolor{black}{In contrast to their quantum counterpart,} they can start with zero initial energy. (b) Sample quantum trajectories, starting in $\ket{0}$, under stroboscopic projective measurement at increment $\theta$.  The trajectories are superpositions of energy eigenstates whose mean energy is shown with a solid line and whose standard deviation is represented by the shading.  \textcolor{black}{Note that the trajectory indicated by the arrow has an average energy larger than $E_B$, but the measurement at $t=5\theta$ results in \textit{survival} and the wavefunction is projected back into the \textit{surviving domain}.}}
    \label{fig:fpt_ILLUSTRATION}
\end{figure}

There are important differences between the classical and the quantum FPTD problem.  In the classical case\textcolor{black}{,} FPTDs \textcolor{black}{are} only defined for stochastic processes. A deterministic process would have a delta-function FPTD. \textcolor{black}{However,} in the quantum case, measurements themselves introduce randomness. 
Thus, even a unitarily evolving quantum system with interspersed measurements has a non-trivial distribution of first-passage times.  
\textcolor{black}{In this letter we focus on the quantized motion of a harmonic oscillator driven by additive noise, although the measurement protocol developed here can be used for any other time evolution.  The first-passage time is defined as the earliest time at which the ion is \textit{measured} to have an energy greater than or equal to a certain barrier \(E_B = \hbar\omega(N_B + \frac{1}{2})\), after initialization in the ground state \(\lvert 0 \rangle\) and under projective measurements performed at fixed intervals \(\theta\). 
% We develop a novel composite laser pulse to experimentally implement single-shot projective measurements to measure escape.  
Fig.~\ref{fig:fpt_ILLUSTRATION} presents sample trajectories (generated using the Monte Carlo wavefunction method~\cite{Molmer:93}) in the energy basis that illustrate key conceptual differences between quantum and classical descriptions of this problem.  In particular, a quantum trajectory can be in a superposition of states both inside and outside the surviving domain, in which case a measurement alters the wavefunction—a fundamentally quantum effect.  In the case that the measurement results in survival, the energy of the trajectory is decreased.  Furthermore, a trajectory whose mean energy is less than the barrier energy may still be detected as having escaped the surviving domain.  In this model the noisy diffusion along the energy ladder is caused by both the additive noise and the quantum projective measurement.}  
Additional quantum properties such as the (anti-) Zeno effect~\cite{Brownian_Motion_Zeno_anti_zeno_PRL} make general Quantum-First-Passage-Time Distributions (QFPTDs) richer than their classical counterpart, and place them at the interface of quantum and classical mechanics.  We note that the classical analogue of this problem is \textcolor{black}{both} non-trivial and subtle due to the multi-dimensional boundary condition~\cite{Hanggi_classical_FPT}. 

Measurements at regular intervals (stroboscopic) have become an accepted method of interpreting the problem of QFPTDs~\cite{Kessler_PRL}, although measurements at random intervals~\cite{Kessler_first_detection_random} and continuous measurements of open quantum systems~\cite{kewmingFPT} have also been investigated.  Since time is not a self-adjoint operator, QFPTDs are ambiguously defined in the continuous measurement limit~\cite{Friedman_PRE, Stovneng_quantum_time, Leavens_quantum_time}.  This ambiguity highlights the need for experimental validation of newly obtained theoretical results—among them, the exotic predictions of spin-dependent arrival times~\cite{das_quantum_time_Nature, das_quantum_time_PRA, das_quantum_time_PRA_II}.  

\textcolor{black}{Quantum walk search algorithms~\cite{quantum_walk_search_1, quantum_walk_search_2, quantum_walk_search_3, quantum_walk_search_4, quantum_walk_search_5} are naturally understood as quantum-first-passage processes~\cite{quantum_walk_search_6, quantum_walk_search_7, Friedman_PRE, Kessler_PRL, Yin_Barkai, Kessler_first_detection_random, BarkaiKesslerQuantumDarkState, BarkaiQFPTD, DharProjectiveQFPTD, Dhar_2015, Krovi_QFPT, Entanglement_QFPT, Experimental_QFPT, kewmingFPT, Gambassi_2025}.}  The development of experimental QFPTD methods has potential to further our understanding of these analytically challenging algorithms with known exponential speedup and support their possible implementation~\cite{quantum_walk_search_7, quantum_walk_search_8_experiment, quantum_walk_search_9_experiment, quantum_walk_search_10_experiment}.  
\textcolor{black}{For example, by using the methods developed here, so-called concurrent measurements which can check the outcome of a quantum walk search while preserving the relevant quantum coherences may be implemented.}  We also expect that QFPTDs can be used for precision measurement as repeatedly monitored quantum systems offer sensing advantages, notably through quantum hindsight effects~\cite{trajectory_sensing_1, trajectory_sensing_2, trajectory_sensing_3}.  

Trapped-atomic ions are an ideal platform for experimental QFPTD measurements. Their prominence in quantum information processing has led to the development of a variety of coherent~\cite{Wineland_Monroe_RevModPhys} and incoherent~\cite{Turchette_engineered_reservoir} control techniques that can be leveraged to simulate a broad variety of QFPTDs.  \textcolor{black}{Furthermore, by coupling the ion to different engineered reservoirs, variations of our QFPTD measurement technique can be used to test dynamical emergence~\cite{Ashida02020}.  In this letter, we measure QFPTDs of a motional degree of freedom of a trapped ion driven by electric-field noise.  We develop a novel widely applicable projective measurement protocol to measure whether or not an energy threshold is exceeded.  We measure QFPTDs for various energy barriers and measurement intervals, and find agreement with theoretical predictions.}

\begin{figure}
    \centering
    \includegraphics[width=1\linewidth]{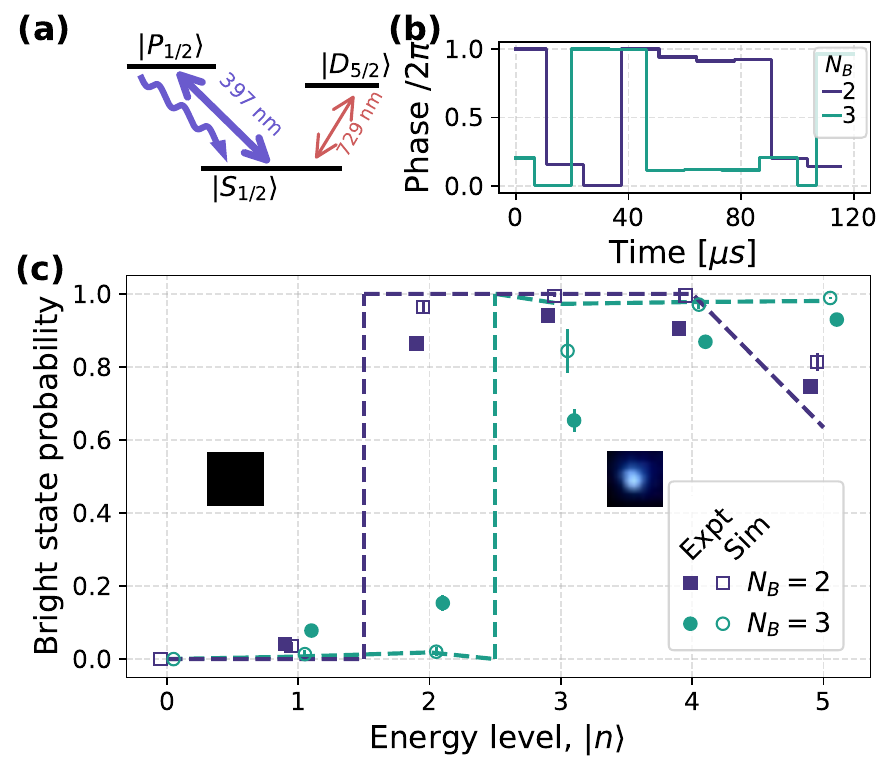}
    \caption{(a) Simplified $^{40}$Ca$^+$ level diagram.  Internal state detection is performed using the dipole-allowed $397$~nm transition, which scatters photons rapidly if in the $\ket{S_{1/2}}$ state, and is dark if in the $\ket{D_{5/2}}$ state.  The composite-phase pulse sequence is done on the quadrupole $729$~nm transition, which hosts the internal qubit.  (b) Composite laser pulse sequence for $N_B = 2,3$.  (c) Step-like excitation profile of the composite phase pulses with thresholds $N_B = 2,3$. The dashed lines are the simulated profile of the step pulse sequence.  The solid points are the experimental results, with $1\sigma$ error bars.  The empty points show the simulated mean excitation (with $1\sigma$ error) given the measured Rabi frequency noise.  For many points, the error bars are are too small to be seen.  \textcolor{black}{We perform about 400 experimental trials for each point.}  See the SM~\cite{supplemental} for detailed error analysis.  We show pictures of the single ion when in the dark state on the left, and in the bright state on the right.}
    \label{fig:step}
\end{figure}

\textcolor{black}{\textit{Projective measurement protocol}}---The \textit{surviving} domain is defined in terms of the energy eigenstates $\ket{n}$ of the unperturbed oscillator: $\{\ket{0},\ket{1},\ket{2}, ..., \ket{N_B-1}\}$.  We define a projective measurement with two possible outcomes (`survival' and `absorption'), whose projectors $\mathbb{P}^\text{S} = \sum_{n=0}^{N_B-1} \ket{n}\bra{n}$ and $\mathbb{P}^\text{A} = \sum_{n=N_B}^\infty \ket{n}\bra{n}$ correspond to the \textit{surviving} and \textit{absorbing} domains, respectively.  We dub the measurement `step pulse' as it effectively forms a quantum-low-pass filter for the energy eigenstates.  \textcolor{black}{In order to implement this measurement on a trapped $^{40}$Ca$^+$ ion, whose internal states are shown in Fig.~\ref{fig:step}a, we use a novel composite-phase laser pulse sequence shown in Fig.~\ref{fig:step}b}.  Composite pulses, first used in NMR spectroscopy, have been used in a variety of contexts to cancel errors (such as the famous \textcolor{black}{Solovay-Kitaev} SK1 and \textcolor{black}{Wimperis broadband} BB1 sequences~\cite{SK1_PhysRevA.70.052318, BB1_1_WIMPERIS1991199, BB1_2_WIMPERIS1994221}), and more recently to detect occupation of harmonic oscillator states~\cite{mallweger2024motional}.  Detuning a $729$~nm laser by the motional frequency above the $\ket{S_{1/2}}\leftrightarrow\ket{D_{5/2}}$ transition gives rise to an anti-Jaynes-Cummings Hamiltonian. This interaction, known as the first blue sideband (BSB), couples the $\ket{S_{1/2}}\otimes \ket{n}$ and $\ket{D_{5/2}}\otimes\ket{n+1}$ states with $n$-dependent coupling strength~\cite{Wineland_Monroe_RevModPhys}.
% {\color{red}\begin{equation}\label{eq:coupling_strengths}
%         \Omega_{n, n+1} = \Omega_{00} e^{\frac{-\eta^2}{2}}\eta \sqrt{\frac{n!}{(n+1)!}} \mathcal{L}_n^{1}(\eta^2).
% \end{equation}}
We exploit the coupling-strength dependence on $n$ to perform an effective $\pi-$pulse on the internal state for $\ket{n \geq N_B}$ \textcolor{black}{(leaving the internal state unaffected for $\ket{n < N_B}$)} using a series of BSB pulses of different relative phases and durations, which we find using numerical optimization.  We show a plot of the expected excitation probability as a function of the motional quantum number for different barriers in Fig.~\ref{fig:step}c, from which the step-like nature of the pulse is clear.  \textcolor{black}{A subsequent projective measurement of the internal state completes the implementation of the projective motional state measurement.}  
In principle, with pulses of sufficient duration and complexity, this conditional excitation can be performed with near-unit fidelity.  However, due to the limited coherence time of the ion and control errors, there is, in practice, a trade-off between the increase in fidelity gained by increasing the duration (and complexity) of the composite pulse, and the effects of decoherence which get worse with increasing time.  See the Supplementary Material (SM)~\cite{supplemental} for more details on the step pulse and these trade-offs.   

In order to measure a QFPTD, we perform repeated trials and aggregate the first-passage times.  For stroboscopic measurements with fixed interval $\theta$, and where $P^{\text{S}(\text{A})}(i\theta)$ is the probability of obtaining a \textit{survival} (\textit{absorption}) outcome from the $i^\text{th}$ measurement given that all prior measurements resulted in \textit{survival}, the QFPTD is given by
\begin{equation}\label{eq:def_fptd}
P^\text{FPT}(k\theta) = \prod_{i=1}^{k-1}P^{\text{S}}(i\theta) \cdot P^{\text{A}}(k\theta).    
\end{equation}  The first passage occurs when the first \textit{absorption} measurement is obtained.  Thus, a single trial proceeds as follows: the ion is ground state cooled using resolved-sideband cooling~\cite{Wineland_Monroe_RevModPhys}, and the internal state is \textcolor{black}{optically} pumped to $\ket{D_{5/2}}$. Thus each trial starts with the ion in
\begin{equation}
\ket{\psi(t=0)}  = \ket{D_{5/2}}\otimes \ket{0}.
\end{equation}
Then, in the quantum trajectories picture, \textcolor{black}{the noisy electric field $\xi(t)$ causes a motional wavefunction $\ket{\psi(t)}$ to evolve for a time $\theta$ according to
$\ket{\psi(t)} \rightarrow \ket{\psi(t+\theta)} = [I \otimes\mathcal{D}(\alpha)] \ket{\psi(t)}$ where
$\alpha = ie / (\sqrt{2m\omega \hbar})\int_t^{t+\theta}dt' \xi(t')e^{i\omega t'}$
and where $\mathcal{D}(\alpha)$ is the displacement operator~\cite{James_alpha_kicks_1998, richerme_2024measurementinducedheatingtrappedions}.  Thus, in the absence of measurements, an initially coherent state is expected to remain coherent under the influence of the noise.  Therefore, before the first measurement, the wavefunction is}
\begin{equation}
\ket{\psi(\theta^-)}  = \ket{D_{5/2}}\otimes \bigg(\sum_{n=0}^{\infty} d_n(\theta)\ket{n}\bigg),
\end{equation}
where $d_n$ are the coefficients of the randomly displaced motional wavefunction.  We then apply the composite-phase pulse \textcolor{black}{with the 729~nm laser} that flips the internal state if the motional state is $\ket{n \geq N_B}$, and leaves the internal state unaffected if the motional state is $\ket{n < N_B}$

\begin{equation}
\begin{split}
\ket{\psi(\theta^+)} &= \ket{D_{5/2}}\otimes \bigg(\sum_{n=0}^{N_B-1} d_n(\theta)\ket{n}\bigg) \\
&\hspace{30pt} + \ket{S_{1/2}}\otimes \bigg(\sum_{N_B}^\infty d_n(\theta)\ket{n}\bigg). \\
\end{split}
\end{equation}
After the composite pulse, the ion is in an entangled state between internal energy levels and the motion.  A subsequent measurement of the ion internal state using state-dependent resonance fluorescence \textcolor{black}{at 397~nm} has two possible outcomes (Fig.~\ref{fig:step}a): $\ket{S_{1/2}}$ in which case the ion fluoresces and photons are rapidly emitted (so-called `bright' state), or it can collapse to the $\ket{D_{5/2}}$ state and emit no photons (so-called `dark' state).  Camera images of the ion in both bright and dark states are shown in Fig.~\ref{fig:step}c.  This measurement of the internal state completes the measurement and projects the motional state either into the \textit{surviving} or \textit{absorbing} domain.
% with projectors $\mathbb{P}^\text{S}$ and $\mathbb{P}^\text{A}$. 
If a \textit{survival} outcome is obtained, \textcolor{black}{the ion is in state
\begin{equation}
    \ket{\psi(\theta^+)}_\text{Survival}  = \ket{D_{5/2}}\otimes \bigg(\frac{1}{\sqrt{P^\text{S}(\theta)}}\sum_{n=0}^{N_B-1} d_n(\theta)\ket{n}\bigg),
\end{equation}
}and the trial continues for another time interval $\theta$ and the step pulse is re-applied.  This sequence is repeated until a bright measurement is obtained, after which the trial is terminated.  \textcolor{black}{We designed this measurement technique such that in the case that a dark state is obtained, a faithful quantum projective measurement has occurred.  However, if a bright state is obtained, the motional state is altered due to to photon recoil.  This has no impact on the FPTD since the trial is ended at that point.  These types of two-step measurement protocols where a first `logic' pulse maps the motional state onto the internal state which is subsequently projectively measured are known to produce faithful quantum measurements of the motional state in the case that a dark outcome is obtained~\cite{Fluhmann2019}.}

\textit{QFPTD predictions}---Having established the QFPTD experimental protocol, we now present a few theoretical predictions.  In this experiment, the ion is coupled to the natural, high-temperature amplitude reservoir from noisy electric fields in the environment~\cite{Turchette_engineered_reservoir}.  This electric-field noise causes an initially ground-state cooled ion to heat and evolve to a thermal state of average occupation number $\bar n$.  We independently measure the heating rate $\Dot{\bar n}$ at the motional frequency $\omega$ through the standard sideband asymmetry technique. We then relate the heating rate to the noise spectral density using $\Dot{\bar n}(\omega) = (e^2/4m\omega\hbar) S_\xi(\omega) $~\cite{Brownnutt}.  We note that, in quantum mechanics, we distinguish between closed systems, which evolve unitarily, and open systems, which do not.  The QFPTD measurement scheme developed in this letter can be used in both cases.  
\begin{figure}
    \centering
    \includegraphics[width=1\linewidth]{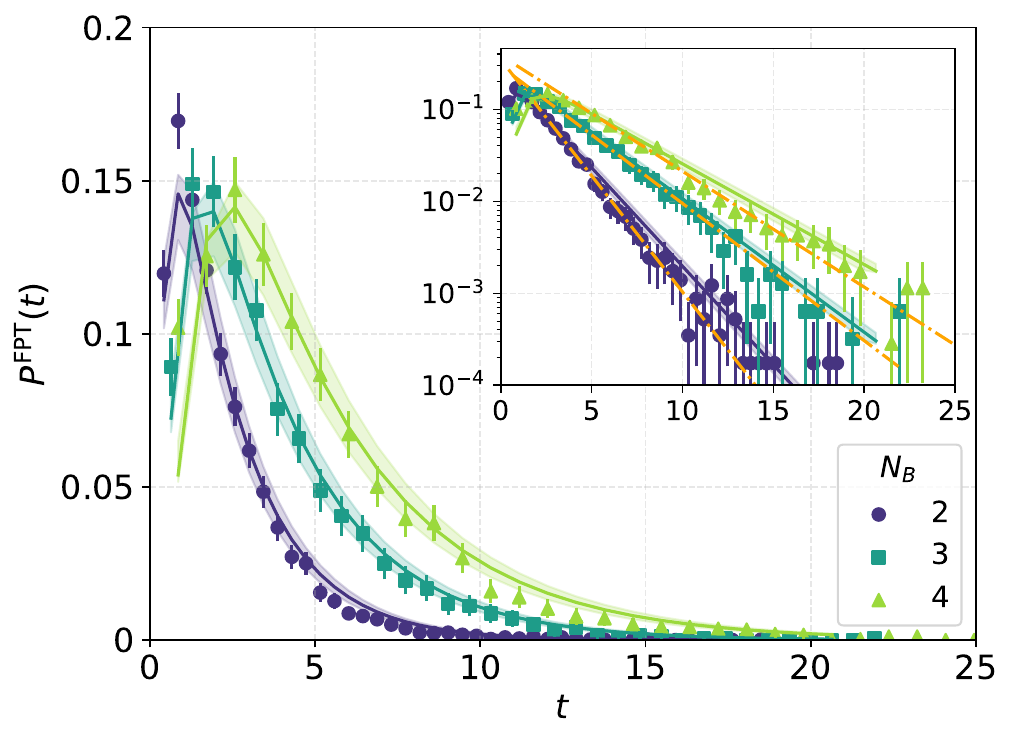}
    \caption{Experimentally measured QFPTDs for $N_B=2,3,4$ with intervals $\theta = 0.43, 0.645, 0.86$ \textcolor{black}{(corresponding to 5, 7.5 and 10ms, respectively)}, and with 6900, \textcolor{black}{3650}, and 4250 trials, respectively.  The points (with $1\sigma$ error bars) are data and the corresponding lines represent theoretical predictions of our model \textcolor{black}{obtained by numerically solving (\ref{eq:def_fptd}) and (\ref{eq:renewal})}.  See the SM~\cite{supplemental} for more details on error analysis.  The shaded error region around the lines represents the $1\sigma$ uncertainty of the measured heating  $\Dot{\Bar{n}}$. The x-axis is dimensionless time $t$, which is related to real time $t'$ by $t = \Dot{\Bar{n}} t'$.  In the inset, the same data are shown with overlaid orange dashed lines which are fits of the measured distribution tails to an exponential.}
    \label{fig:experimental_FPT}
\end{figure}

\begin{figure}[t]
    \centering
    \includegraphics[width=1\linewidth]{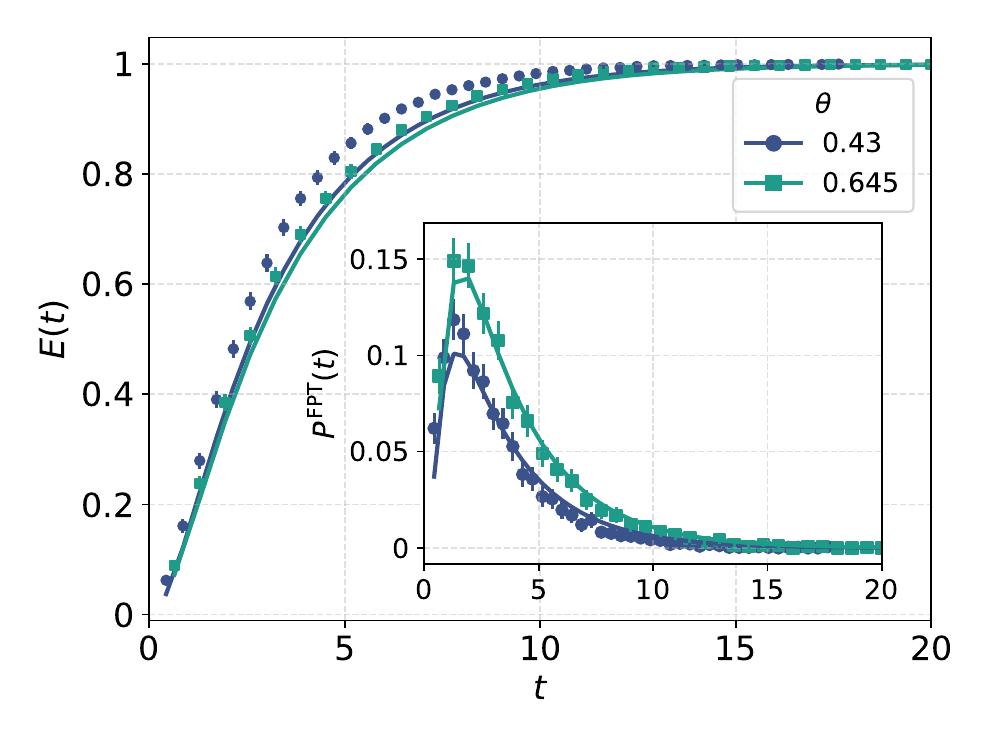}
    \caption{Escape probability $E(t)$ and QFPTD $P^\text{FPT}(t)$ (inset) for $N_B=3$ with $\theta = 0.43, 0.645$\textcolor{black}{, and with 3800 and 3650 trials respectively}.  The experimental data are shown using the scattered points, and the theoretical predictions are shown using solid lines.  The error bars represent $1\sigma$ uncertainties.  For the escape probability, they reflect the negative covariance between the probability estimators in the QFPTD, as the underlying statistics follow a multinomial distribution.  See the SM~\cite{supplemental} for details on error bars.}
    \label{fig:double-speed}
\end{figure}

The evolution of the motional state due to the electric-field noise in between measurements can be described using either quantum trajectory theory, or a quantum master equation.  For the density matrix
$\rho$, the master equation is~\cite{Turchette_engineered_reservoir}
\begin{equation}\label{eq:master_eq}
    \Dot{\rho}    
= \frac{\Dot{\bar n}}{2} (2a\rho a^\dagger - a^\dagger a \rho - \rho a^\dagger a + 2a^\dagger \rho a -a a^\dagger \rho -\rho a a^\dagger)
\end{equation}
where $a^\dagger (a)$ is the motional state creation (annihilation) operator.  \textcolor{black}{We also write (\ref{eq:master_eq}) as $\dot \rho = \mathcal{M}\rho$.}
This evolution closely resembles a well-known classical analogue: the classical undamped harmonic oscillator driven by additive noise: $
    \Ddot{x} + \omega^2x = e\xi(t)/m$ 
where $m$ is the mass of the ion\textcolor{black}{~\cite{Gardiner_quantum_noise, Bruerer_and_pretrucione, HArty:2013}}.  For convenience, we henceforth adopt dimensionless time $t$, such that $\Dot{\bar n} = 1$, \textcolor{black}{and use the independently measured value of $\Dot{\Bar{n}}$ to relate experimental and theoretical timescales.  }
% \textcolor{black}{We use this definition of dimensionless time as it makes the mean first-passage time similar to $N_B$ (exactly for $\theta=0$, and approximately for finite-$\theta$).  }
\textcolor{black}{The normalized density matrix representing the surviving states immediately after the $n$-th measurement at $t=n\theta$ is given by a renewal equation
\begin{equation}\label{eq:renewal}
    \rho\big(t=n\theta^+\big)_\text{Surviving} = \frac{\Big[\mathbb{P}^\text{S} e^{\theta\mathcal{M}}\Big]^{n} \rho(t=0)\Big[ \mathbb{P}^\text{S}\Big]^n}{P^S(\theta)P^S(2\theta)...P^S(n\theta)}.
\end{equation}
from which the QFPTD is calculated.  The diffusion along the energy ladder is then due to both the noisy evolution and quantum projection noise.  Furthermore, we note that (\ref{eq:master_eq}) may admit non-classical initial conditions~\cite{Gardiner_quantum_noise}, and due to the projective measurements, the surviving states in (\ref{eq:renewal}) become non-classical~\footnote{Manuscript in preparation}.}  \textcolor{black}{We numerically solve (\ref{eq:renewal}) to obtain the QFPTD and show the results in Fig.~\ref{fig:experimental_FPT} along side the experimental results.}  
For finite~$\theta$, we predict a reduction of the mean first-passage time for decreasing $\theta$.  This effect is reflected in the escape probability $E(t; \theta) = \sum_{i=1}^{t/\theta}P^\text{FPT}(i\theta)$ at time $t$, which is larger for smaller $\theta$: i.e. $E(t; \theta_1) < E(t; \theta_2)$ if $\theta_1 > \theta_2$.  
This enhancement is shown in the theoretically predicted curves in Fig.~\ref{fig:double-speed}.  Although it appears to be an anti-Zeno effect, the enhancement of the escape probability arises because faster probing detects an escape sooner.  This process is analogous to evaporative cooling of atom clouds (in particular to the RF knife technique~\cite{Evaporative_Cooling_RevModPhys.84.175}), where in this case, surviving trajectories have lower average energy than if non-selective measurements had been performed.  Furthermore, due to the nonclassical quantization of energy, for $N_B=1$ the QFPTD is a pure exponential (shown in the SM~\cite{supplemental}) without an initial ballistic part seen for $N_B\geq 2$, as each measurement resulting in survival projects the \textcolor{black}{motional state} back into $\ket{0}$, and thus \textcolor{black}{the first-passage process follows a geometric distribution as the probability of measuring an escape is fixed.}
% \sout{the trajectory begins in its statistical steady state}.  However, for $N_B\geq 2$ the statistical steady state is only achieved around $t\gtrsim N_B$, at which point the QFPTD transitions from the ballistic to the diffusive regime.  
This is \textcolor{black}{described} in the SM~\cite{supplemental}.

\textit{Experimental results}---\textcolor{black}{We perform these experiments in a cryogenic linear Paul trap whose heating rate, measured by sideband asymmetry, is $\dot {\bar n} = 86\pm 8$ q/s.  We use a radial mode with Lamb-Dicke parameter $\eta = 0.055$ and carrier Rabi frequency $\Omega_{00}\sim 2\pi\times 300$ kHz.}  We validate the selective excitation profile of the step pulse by creating number states of motion $\ket{n}$, applying the step pulse, and then measuring the internal state excitation probability.  
The number states are created by first ground state cooling the ion to $\ket{S, 0}$ and then using a sequence of BSB $\pi$-pulses of the appropriate duration~\cite{Wineland_Monroe_RevModPhys}.  The results of the test and the expected excitation profile are shown in Fig.~\ref{fig:step}c.  
The overwhelming majority of the errors in the step pulse measurement sequence come from the composite-phase pulse, and not the subsequent state-dependent fluorescence which is performed with error $<10^{-3}$.
\textcolor{black}{We mainly attribute errors in the step pulse to laser intensity noise caused by vibrations from the cryostat motor in the experimental assembly, leading to beam-pointing instability.  This causes the Rabi frequency to drift and fluctuate.  See the SM~\cite{supplemental} for a more detailed analysis of the Rabi frequency noise.}
In Fig.~\ref{fig:step}c, we show simulated results with Rabi frequency noise (estimated from independent measurements) alongside experimental data.
We also have contributions from imperfect ground state cooling (which results in lower number state creation fidelity), and smaller contributions from internal state decoherence ($T_2\sim580\hspace{2pt}\mu s$) due to slow B-field fluctuations, and motional state decoherence ($T_2\sim280\hspace{2pt}\mu s$). See the SM~\cite{supplemental} for a more detailed error analysis and for details of the setup and control system.

Experimental QFPTD measurement results for $N_B = 2,3,4$ are shown in Fig.~\ref{fig:experimental_FPT}.  
The experimental results—notably their long-time behavior—show good agreement with the theoretical predictions and are consistent with an exponential decay at long times.  
As $N_B$ increases, the ballistic portion is longer, in good agreement with theory.  Pulse errors tend to increase the probability of early detection; thereby shifting the QFPTD forward in time.  This effect also contributes to an exponential tail with a shorter characteristic time than that of the corresponding theoretical prediction.  In order to compare the effect of moving $N_B$ without becoming dominated by step pulse errors, in Fig.~\ref{fig:experimental_FPT} we scale $\theta$ proportionally to $N_B$.  

In a separate experiment we explore a critical aspect of stroboscopic QFPTDs: the dependence of the escape probability on $\theta$.  We compare QFPTDs with $N_B=3$ and different $\theta$ in Fig.~\ref{fig:double-speed}.  Since the QFPTDs shown here are probabilities and not probability densities, direct comparison of the QFPTDs for different values of $\theta$ is done using the escape probability.  The experimental results are consistent with the presence of an enhancement of the escape probability for smaller $\theta$.  However, with current step pulse error rates, these data do not conclusively demonstrate it.  This enhancement is a consequence of the repeated measurements within each run \textcolor{black}{which modify the trajectory}.  If instead measurements were performed once in a run, then an `escape' probability could be reconstructed, but it would be insensitive to $\theta$.  

We have tacitly assumed that there is no spontaneous emission of the $\ket{D_{5/2}}$ state (lifetime 1.2s~\cite{KreuterSpontMish}) between stroboscopic measurements.  The duration of these experiments is sufficiently short that errors due to spontaneous emission do not dominate.   While this poses a limitation to directly measuring QFPTDs which go out to long times, this problem can be circumvented by using either longer-lived optical qubits, such as Barium~\cite{Barium_spont_mish}, or ground-state qubits with practically infinite lifetimes~\cite{Kang_Mingyu_practically_infinite_lifetime}.  Furthermore, given knowledge of the excited state lifetime, the true QFPTD can be reconstructed from the observed QFPTD which has been modified by spontaneous emission.   

\textit{Conclusions}---We develop and validate a novel way to use the motion of a trapped ion to measure QFPTDs.  We use this method to measure experimentally the QFPTD of the ion energy when it is coupled to a high-temperature amplitude reservoir, and show good agreement with theory.  
The step pulse measurements developed here can potentially be used for bosonic state engineering and simulation \cite{katz_bosonic, Brown_Bosonic}.  Further engineering of bespoke measurement operators raises the possibility of studying the properties of QFPTDs with more exotic surviving domains, such as quantum recurrence times~\cite{KissQFPTDPolya} and winding numbers~\cite{winding_1, winding_2}.  
Complex interaction graphs for applications such as quantum walk search algorithms can be implemented using more co-trapped ions thereby increasing the number of motional and spin degrees of freedom \cite{Monroe_qsim, kihwan_qsim, katz_bosonic}.  Using more trapped ions, the role of entanglement in QFPTDs could be investigated, which remains largely unexplored.

\textit{Acknowledgments}---We are grateful to Jude Alnas for assistance with ARTIQ control.
This work was funded by NSF under QLCI: Center for Robust Quantum Simulation OMA-2120757. JR is supported by the Goshaw Fellowship.  This manuscript was edited in part at the Aspen Center for Physics, which is supported by National Science Foundation grant PHY-2210452.

\textit{Data availability}---The data that support the findings of this article are openly available~\cite{open_data}.

\bibliography{bibliography}
\end{document}

\preprint{APS/123-QED}

\title{\textbf{Supplementary Material: Experimental measurement of quantum-first-passage-time distributions} 
}% 

\author{Joseph M. Ryan}
\email{Contact author: joseph.ryan@duke.edu}
\author{Simon Gorbaty}
\author{Thomas J. Kessler}
\author{Mitchell G. Peaks}
\author{Stephen W. Teitsworth}
\author{Crystal Noel}
% \author{Simon Gorbaty}%
%  \email{Contact author: Second.Author@institution.edu}
\affiliation{Duke Quantum Center, Department of Electrical and Computer Engineering and Department of Physics, Duke University, Durham, NC 27708, USA}%

\date{\today}% It is always \today, today,

\maketitle

\onecolumngrid
% \begin{center}
%     \textbf{\large Supplementary Material: Experimental measurement of quantum-first-passage-time distributions}
% \end{center}
% \twocolumngrid

\appendix

\section{Experimental setup}
We perform these experiments in a cryogenic linear Paul trap, with radial mode frequencies of 3.1 MHz and 3.3 MHz.  The trap axes are rotated such that the 729 nm beam used for the coherent composite-phase pulse is nearly collinear with the lower frequency (slow) mode (Lamb-Dicke parameter $\eta = 0.055$ and carrier Rabi frequency $\Omega_{00} \sim 2\pi\times300$~kHz), which we use to perform these measurements.  We perform sideband cooling on both radial modes and typically achieve $\bar n = 0.07$ on both modes.  The heating rate of the slow modes—which determines the dimensionless time—is independently measured by sideband asymmetry to be $\Dot{\bar n} = 86 \pm 8 $~q/s.  

\begin{figure}[b]
    \centering
    \includegraphics[width=0.5\linewidth]{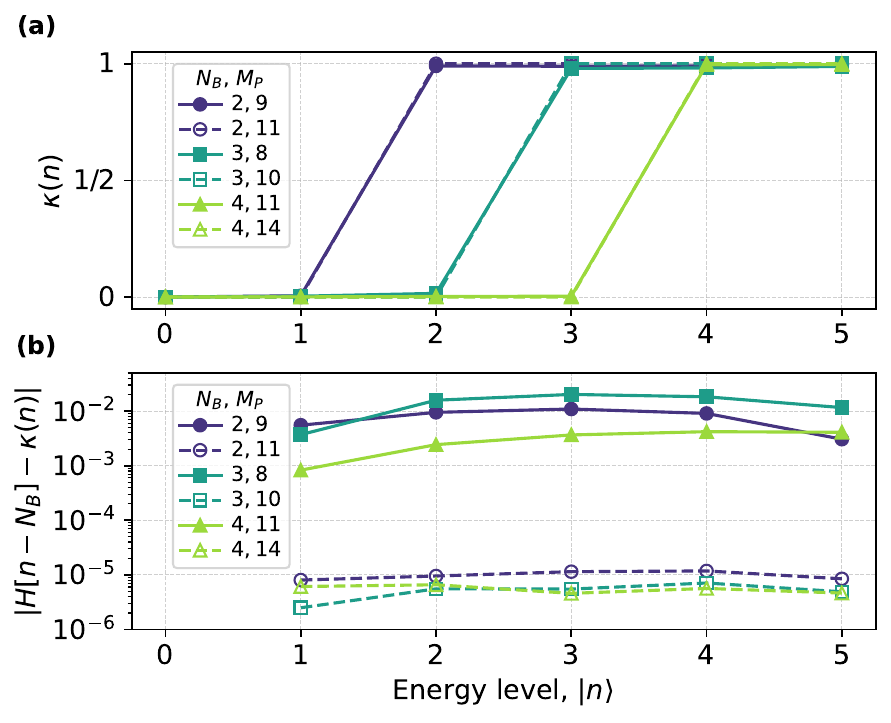}
    \caption{(a) Rotation probability $\kappa(n)$.  (b) The rotation error $|H[n-N_B] - \kappa(n)|$ decreases with increasing $M_P$.  $H[i]$ is the discrete Heaviside step function, defined as $H[i] = 0$ for $i<0$, and $H[i] = 1$ for $i\geq0$.}
    \label{fig:log_step}
\end{figure}

\section{Composite-phase step pulse sequence}\label{app:pulseSequence}
Tuning the 729nm laser beam to the first blue sideband (BSB) gives rise to an anti-Jaynes-Cummings interaction \cite{Wineland_Monroe_RevModPhys}
$$H_I = \frac{\hbar}{2}\Omega_{0,0}\eta (a^\dagger\sigma^+ e^{i\phi} +a \sigma^- e^{-i\phi})$$where $\phi$ is the laser phase and $\sigma_{\pm}=(\sigma_x\pm i\sigma_y)/2$ is the spin raising and lowering operator corresponding to the two-level system formed by the $\ket{S}$ and $\ket{D}$ levels. This interaction couples the manifolds of states $\ket{S, n}\leftrightarrow\ket{D,n+1}$ with coupling strength $\Omega_{n, n+1}$ \cite{Wineland_Monroe_RevModPhys}
\begin{equation*}\label{eq:coupling_strengths}
        \Omega_{n, n+1} = \Omega_{00} e^{\frac{-\eta^2}{2}}\eta \sqrt{\frac{n!}{(n+1)!}} \mathcal{L}_n^{1}(\eta^2)
\end{equation*}
where $\mathcal{L}_n^{1}(x)$ is the first generalized Laguerre polynomial.  We exploit the coupling-strength dependence on $n$ to perform an effective $\pi-$pulse on the internal state if $\ket{n \geq N_B}$.  This step pulse is formed by applying a sequence of $M_P$ pulses with phases $\phi_i$ for durations $t_i$, which we find using a Trust Region Reflective numerical optimizer \cite{scipy}.  By increasing the number of pulses $M_P$, and their duration, step pulses of increasing fidelity can be made.  The phases and durations of the step pulses are shown in Table \ref{tab:example}.  The durations are shown in terms of the Rabi frequency $\Omega_{00}$ of the carrier transition $\ket{S,0}\leftrightarrow\ket{D,0}$.  We show the rotation probability $\kappa (n)$ in Fig.~\ref{fig:log_step}.  If the internal state begins in $\ket{D_{5/2}}\otimes \ket{n}$, the probability of a flip to $\ket{S_{1/2}}\otimes\ket{n-1}$ is $\kappa(n)$.  To visualize deviation from a perfect discrete Heaviside step function $H[n]$, we also plot $|H[n-N_B] - \kappa(n)|$, where $H[i] = 0$ for $i<0$, and $H[i] = 1$ for $i\geq0$.

\section{Rayleigh intensity noise}\label{app:Rayleigh_noise}

\begin{figure}
    \centering
    \includegraphics[width=0.5\linewidth]{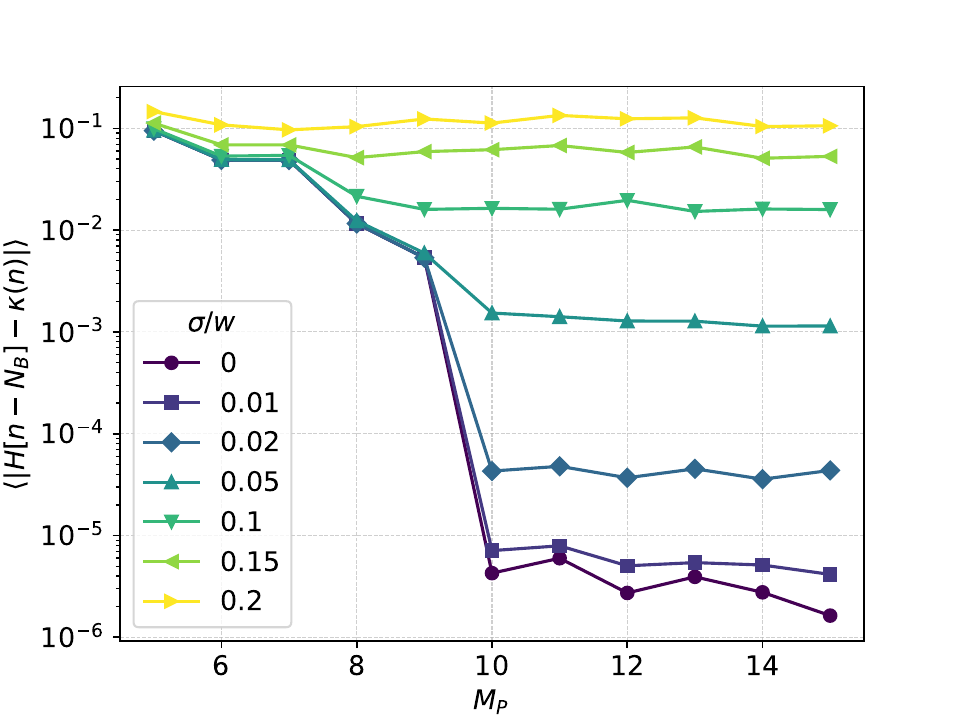}
    \caption{Effect of Rayleigh intensity noise strength $\sigma/w$ on the mean rotation error $\langle|H[n-N_B] - \kappa(n)|\rangle$, for $N_B=3$, and different number of pulses $M_P$.  In our experiment, $\sigma / w = 0.13$.}
    \label{fig:tradeoff}
\end{figure}

The primary source of error when applying the composite pulse sequence comes from intensity fluctuations of the 729 nm beam.  In particular, due to cryostat motor vibrations, we have substantial beam-pointing instability.  The beam is Gaussian with waist $w$.  Previous measurements of the vibration frequency \cite{Pagano_2019} suggest that it is dominated by slow noise.  Thus, we assume that the beam pointing is constant over the duration of one shot ($\sim100~\mu$s).  We do not assume that the laser intensity fluctuations at the ion position are Gaussian.  Rather, we assume that the laser beam center's fluctuations are Gaussian.  We assume that the beam's center $x, y$ is noisy: $x,y \sim N(0, \sigma^2), N(0,\sigma^2)$.  Then, the distance $r$ from the center is distributed as:
$$r = \sqrt{x^2+y^2} \sim \text{Rayleigh}(\sigma) = \frac{r}{\sigma^2}e^{\frac{-r^2}{2\sigma^2}}.$$ We use the predicted decay envelope of Rabi oscillations under this noise to find $\sigma / w = 0.13$ in our system.  We show the effect of this noise on the mean rotation probability error $\langle|H[n-N_B] - \kappa(n)|\rangle$ averaged over the first 6 energy levels, for different $M_P$ and noise strengths $\sigma / w$ in Fig.~\ref{fig:tradeoff}.

\textcolor{black}{There are numerous other sources of errors that are not taken into account in the simulated mean excitation in Fig.~2 of the main text, which only takes into account measured Rabi frequency noise.  We chose to add only this minimal noise model to the simulation as adding more would obscure its interpretability, and reduce its faithfulness as more noise parameters would be needed.  The measured Rabi frequency noise is the shot-to-shot variation in Rabi frequency over the course of several minutes.  However, the experimental data shown in Fig.~2 of the main text is taken over the course of days.  Over these longer time scales, we observed significant variations in the Rabi frequency noise, as well as in other parameters such as the Doppler cooling and sideband cooling quality.  Furthermore, the simulation model used in Fig.~2 in the main text does not take into account the imperfect ground state cooling or the number state creation fidelity.  Since the number states are created by applying a sequence of BSB $\pi$-pulses, the number state creation fidelity gets worse for larger $\ket{n}$.  While we re-calibrate the Rabi frequency several times per hour, we also observed during the course of experiments that there were times during the day that the experiment was far noisier than others, which was primarily due to fluctuations in the laboratory environment.  }

\begin{table}[t]
    \caption{Pulse sequence for various $N_B$ values, $\eta = 0.05533$}
    \label{tab:example}
    \begin{ruledtabular}
    \begin{tabular}{c 
                    p{2.4cm} 
                    p{2.9cm} 
                    p{2.9cm}}
        $N_B$ & $M_P$ & Phase & Duration \\
        \midrule
        2 & 9
          & (0.867, 0.725, 0.791, 0.960, 0.325, 0.391, 1.870, 1.370, 1.264)$\pi$
          & (4.000, 4.000, 3.904, 4.000, 4.000, 4.000, 3.720, 4.000, 4.000)$/(2\Omega_{00})$ \\
          & 10
          & (1.013, 1.125, 1.028, 0.637, 0.888, 0.397, 0.086, 1.854, 1.664, 1.433)$\pi$
          & (4.000, 4.000, 4.000, 4.000, 4.000, 1.542, 4.000, 3.334, 4.000, 3.149)$/(2\Omega_{00})$ \\
          & 11
          & (1.715, 1.333, 1.520, 1.477, 1.045, 1.152, 0.740, 0.922, 0.763, 0.398, 0.696)$\pi$
          & (3.987, 3.874, 3.394, 3.338, 3.786, 3.568, 3.999, 2.575, 3.862, 4.000, 3.999)$/(2\Omega_{00})$ \\
        \midrule
        3 & 8
          & (1.588, 1.663, 0.000, 1.622, 1.755, 1.508, 1.714, 1.562)$\pi$
          & (6.000, 6.000, 6.000, 4.829, 2.653, 4.479, 5.400, 4.097)$/(2\Omega_{00})$ \\
          & 9
          & (1.107, 0.927, 1.108, 0.986, 1.161, 0.857, 1.136, 1.042, 0.981)$\pi$
          & (4.055, 5.448, 4.218, 3.791, 4.513, 4.907, 4.983, 5.466, 2.659)$/(2\Omega_{00})$ \\
          & 10
          & (1.107, 0.875, 1.321, 1.562, 0.098, 0.302, 1.974, 0.340, 0.720, 1.359)$\pi$
          & (3.677, 5.311, 6.000, 3.712, 6.000, 5.992, 4.828, 5.114, 3.760, 3.498)$/(2\Omega_{00})$ \\
        \midrule
        4 & 10
          & (1.640, 0.006, 1.952, 1.921, 1.784, 1.897, 1.902, 1.982, 1.693, 0.000)$\pi$
          & (2.365, 4.099, 4.716, 1.680, 3.983, 5.863, 4.654, 4.994, 4.455, 5.243)$/(2\Omega_{00})$ \\
          & 11
          & (2.000, 1.841, 1.895, 1.771, 0.000, 0.428, 0.225, 0.581, 0.322, 0.203, 0.166)$\pi$
          & (1.656, 5.097, 5.998, 6.000, 6.000, 5.008, 2.548, 5.621, 4.105, 6.000, 6.000)$/(2\Omega_{00})$ \\
          & 14
          & (1.147, 1.463, 1.757, 0.001, 0.634, 0.750, 0.401, 2.000, 0.036, 1.037, 0.407, 0.306, 0.562, 0.539)$\pi$
          & (5.830, 5.999, 5.643, 3.863, 5.829, 5.114, 3.353, 5.205, 5.755, 0.127, 3.499, 3.799, 5.216, 4.155)$/(2\Omega_{00})$ \\
    \end{tabular}
    \end{ruledtabular}
\end{table}
% \begin{figure}[h]
%     \centering
%     \includegraphics[width=1\linewidth]{Plots/histogram-2.pdf}
%     \caption{P.D.F. of $E/E_0$ for different values of $\sigma/w$.  }
%     \label{fig:histo}
% \end{figure}

% \nocite{*}

\section{FPT sampling error bars}
The experimental measurement of a discrete-time FPTD is a multinomial sampling problem.  Suppose that we perform $n$ trials, and that within each trial we perform $k-1$ step-pulse measurements.  Thus, within each trial, there are $k$ possible outcomes since it is possible for a run not to terminate within the $k-1$ measurements, i.e. the first-passage time $T$ is drawn from $T\in\{T_\theta, T_{2\theta}, T_{3\theta}, ..., T_{(k-1)\theta}, T_{\geq k\theta}\}$.  Suppose that for each outcome $T_i$, we obtain $X_i$ successes, then $T_i\sim\text{Bin}(n, p_i)$.  So, for the FPTDs shown in  Fig.~3 and~4 in the main text, we use the binomial MLE estimator $\Hat{p}_i = X_i/n$ and $\text{Var}(\Hat{p}_i) = p_i(1-p_i)/n$.  In those figures, we show the $1\sigma$ confidence interval.  However, when calculating the escape probability $E$, we must take into account the fact that the $\Hat{p}_i$ are not independent, they are negatively correlated, since they satisfy the normalization constraint, $\sum_i p_i=1$.  Indeed, the covariance matrix off-diagonal terms are: $\text{Cov}(\Hat{p}_i, \Hat{p}_j) = \Hat{p}_i\Hat{p}_j/n$ for $i\neq j$.   Thus, for each $i$, the measured escape probability $\Hat{E}(i\theta) = \mathbf{a}_i\Vec{\Hat{p}}$, where
\begin{equation*}
\mathbf{a}_i = ( \underbrace{1,\,1,\,\dots,\,1}_{i \text{ times}},\ \underbrace{0,\,0,\,\dots,\,0}_{k-i \text{ times}} )
\end{equation*}
and $\Vec{\Hat{p}}$ is a vector of the probability estimators $\Hat{p}_i$.  Then, $\text{Var}(\Hat{E}(i\theta)) = \mathbf{a}_i\text{Cov}(\Hat{p}) \mathbf{a}_i^T$.  This means that the error bars of the escape probability are largest when $E\sim 1/2$, and are small when $E\sim0, 1$.  The error bars shown in Fig.~3 and~4 in the main text only represent sampling error from due to the finite number of trials.  They do not take into account uncertainty in the pulse fidelity.

\section{QFPTD with $N_B$=1 is different from $N_B\geq 2$}

\begin{figure}[h!]
    \centering
    \includegraphics[width=0.5\linewidth]{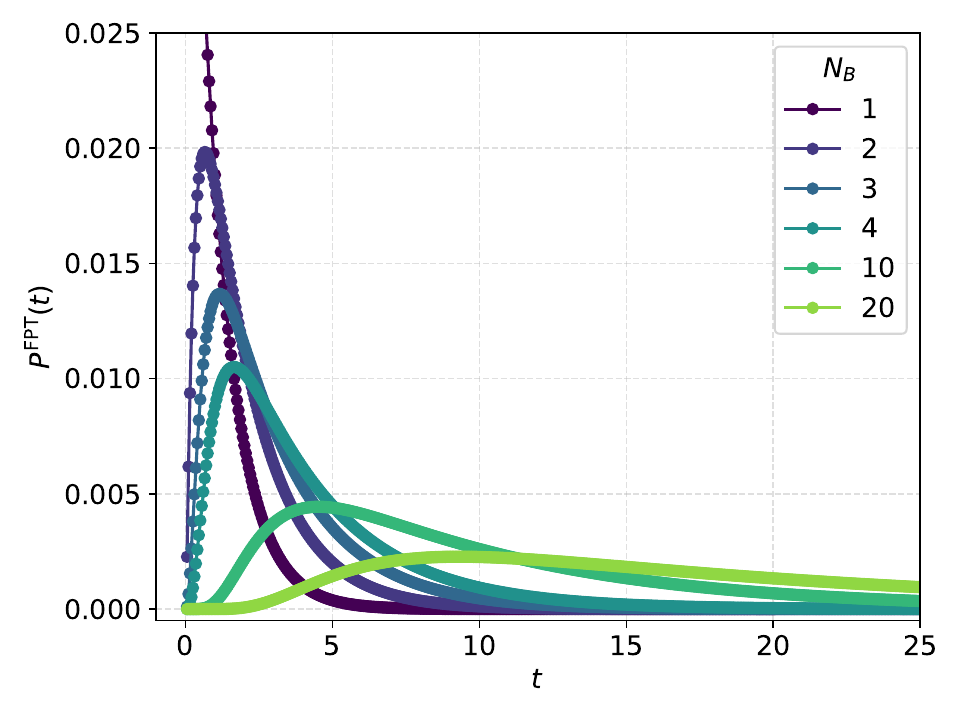}
    \caption{QFPTDs for $\theta = 0.05$ and various $N_B$.  These are calculated using both quantum master equations and quantum trajectories, which yield identical results.  The long-time tails of these QFPTDs are exponential.}
    \label{fig:qfptd_discrete}
\end{figure}

Due to the quantization of energy, the QFPTD for $N_B=1$ is qualitatively different from those where $N_B\geq 2$. In the $N_B=1$ case, it is a pure exponential and does not have the initial ballistic behavior seen in those where $N_B\geq 2$.  In this case, a measurement that results in survival projects the motional wavefunction back into the ground state $\ket{0}$.  Thus, the subsequent measurement has a fixed probability of resulting in absorption.  That probability is the probability that a thermal distribution with $\bar n = \theta$ is not found in $\ket{0}$.  For finite $\theta$, this probability is $p=\theta/(\theta + 1)$.  Thus, the QFPTD for $N_B=1$ is a geometric distribution, where the first-passage time $T\sim\text{Geo}(p)$, where $p=\theta/(\theta + 1)$.  Naturally, in the limit of continuous measurement, the QFPTD for $N_B=1$ is an exponential distribution where the first-passage time $T\sim\text{Exp}(\lambda = 1)$.  QFPTDs for $\theta = 0.05$  are shown in Fig.~\ref{fig:qfptd_discrete}, where the qualitative differences between $N_B=1$ and $N_B\geq 2$ are clear.

\section{Experimental control setup}\label{app:ControlSystem}
Our experiment uses the ARTIQ control environment. Each stroboscopic measurement in the FPT experiment consists of the 729~nm laser composite pulse sequence followed by a measurement  of the ion’s internal state accomplished with resonance fluorescence at 397 nm.  The resonance fluorescence is measured using a PMT, whose TTL output is sent to the ARTIQ controller.  

The phase of the 729~nm pulse is modulated by an AOM, which is controlled with an RF signal from an AD9910 DDS.  The phases are written on the  1024x32 bit RAM on an AD9910.  Using the `polar mode' of the DDS, both the amplitude and phase for one step can be specified inside a single 32-bit address block.  We use the `polar mode' to avoid ringing effects that arise from using the built-in attenuator to control the amplitude.  Due to the delay incurred when loading the pulse sequence onto the RAM with ARTIQ, we load the entire chain of phases and amplitudes for all the attempts prior to the start of a measurement run.  The limited size of the DDS RAM places a limit on the complexity of the composite pulse.

We measure the ion state by shining the 397~nm laser and the 866 re-pumper for $200~\mu$s, and simultaneously counting collected photons on the PMT which are registered as rising edges on a TTL channel.  The rising edge count is stored in a buffer, and the value can then be fetched and the buffer cleared.  However, this operation is slack consuming and performing it in between stroboscopic measurements would necessitate adding an extra delay after each measurement so that the timeline of operations does not fall behind the reference clock.  We bypass the need for this delay by storing all the rising edge counts in the buffers, and making all fetch calls at the end of the shot to ensure no possibility of the reference clock overtaking the timeline.   This limits the number of measurements to 64, which is the number of available buffers.  This ensures that we can maintain the deterministic timeline for the experiment.

\bibliography{bibliography}